\documentclass[%
 aip,
 chaos,%
 amsmath,amssymb,
 reprint,%
]{revtex4-1}

\usepackage{graphicx}
\usepackage{dcolumn}
\usepackage{bm}

\usepackage{color}

\begin{document}

\preprint{AIP/123-QED}
\title{On a simple model that explains inversion of a self-propelled rotor under periodic stop-and-release operations}

\author{Yuki Koyano}
\affiliation{Department of Physics, Graduate School of Science, Tohoku University, 6-3, Aoba, Aramaki, Aoba-ku, Sendai 980-8578, Japan}

\author{Hiroyuki Kitahata}
\affiliation{Department of Physics, Graduate School of Science, Chiba University, 1-33 Yayoi-cho, Inage-ku, Chiba 263-8522, Japan}

\author{Satoshi Nakata}
\affiliation{Graduate School of Integrated Sciences for Life, Hiroshima University, 1-3-1 Kagamiyama, Higashi-Hiroshima 739-8526, Japan}

\author{Jerzy Gorecki}
\email{jgorecki@ichf.edu.pl}
\affiliation{Institute of Physical Chemistry, Polish Academy of Sciences, Kasprzaka 44/52, 01-224 Warsaw, Poland}

\date{\today}

\begin{abstract}
We propose a simple mathematical model that describes the time evolution of a self-propelled object on a liquid surface using such variables as the object location, the surface concentration of active molecules and the hydrodynamic surface flow.
The model is applied to simulate the time evolution of a rotor composed of a polygonal plate with camphor pills at its corners. We have qualitatively reproduced results of experiments, in which the inversion of rotational direction under periodic stop-and-release operations was investigated.
The model correctly describes the probability of the inversion as a function of the duration of the phase when the rotor is stopped.
Moreover, the model allows to introduce the rotor asymmetry unavoidable in real experiments and study its influence on the studied phenomenon.
Our numerical simulations have revealed that the probability of the inversion of rotational direction is determined by the competition among the transport of the camphor molecules by the flow, the intrinsic asymmetry of the rotor, and the noise amplitude.
\end{abstract}


\maketitle
 
\begin{quotation}
There are many solid substances, such as camphor, camphene, and phenanthroline, that develop to the water surface as a molecular layer, evaporate to the air phase and continuously dissolve into the water phase.
This surface layer decreases the water surface tension, and this decrease is local and time dependent.
Inhomogeneities in the surface concentration, resulting from fluctuations, can break the symmetry around a symmetrical object and drive its motion.
There are many interesting examples of systems where the motion of a self-propelled object is strongly coupled with generated hydrodynamic flows.
 Recently, interesting experimental results have been reported~\cite{NakataJPCC}: when a rotor composed of a regular hexagonal plate with camphor pills at its corners is floated on the water surface, it exhibits either clockwise or counterclockwise rotation.
After a forced stop and subsequent release, the rotor starts to move again.
If the stop phase is short enough, the rotational direction is inverted if compared to the rotational direction before the stop operation.
Therefore, the system has a memory of the previous rotational direction.
This information is preserved in the water flow around the rotor and gets lost when the stop phase is long.
To address the mechanism of this inversion phenomenon, a new mathematical model describing time evolution of self-propelled objects and including the flow effects is presented in the paper.

A standard simple model of self-propelled motion treats the time evolution of surface concentration of the active substance, as a reaction-diffusion process.
However, such a model does not describe systems where hydrodynamic flows are directly related to the surface concentration.
Here, we introduce a new model in which hydrodynamic flows and the surface concentration of the active substance are independent variables.
Using the model, we have qualitatively reproduced the experimental results mentioned above.
Moreover, the model allows to introduce the rotor asymmetry unavoidable in real experiments and study its influence on the inversion probability.
Our numerical study has revealed that this probability is determined by the competition among the transport of camphor molecules by the flow, the intrinsic asymmetry of the rotor, and the amplitude of fluctuations.
The model is general and can be adopted to other systems studied experimentally where the coupling between self-propelled motion and hydrodynamics is important.
\end{quotation}

\section{Introduction}

Self-propelled objects have been intensively studied in the last decades~\cite{Ramaswamy2010,Marchetti2013,Bechinger2016,Pimienta2014,Keren}.
Many of them are driven by the gradient of surface tension or interfacial tension in nonequilibrium condition.
They represent an interesting manifestation of nonequilibrium evolution and mimic the motions of living organisms~\cite{Taylor2013,Park2017,Frenkel2017,Frenkel2017,Nagai2005,Hanczyc2014,Lagzi2010,Takanatake2014,Toyota2009,Izri2014,Sumino2005,Cira2015,Domingues1995}.
Symmetric properties of self-propelled objects are important to understand the character of their motion~\cite{Ohta2017}.
The character of motion of an asymmetric object reflects its geometry.
For a symmetric object, a non-zero force can emerge through spontaneous symmetry breaking.
A number of reports focusing on the symmetric properties of self-propelled objects have been published recently~\cite{OhtaOhkuma,Ebata1,Ebata2,KoyanoJCP}.

One of the most studied self-propelled systems is a piece of camphor on a water surface~\cite{Skey,Tomlinson,Rayleigh,NakataLangmuir,Nakata2015,NakataBook}.
The physical mechanism of self-propelled motion can be easily explained.
Camphor molecules are continuously released from the solid particle to the water surface.
They poorly dissolve in bulk water and make a layer on the surface.
The water surface tension is a decreasing function of the camphor surface concentration~\cite{camphor_surface_tension,Karasawa}.
The local changes in camphor surface concentration, resulting from system geometry, flows, and fluctuations in camphor evaporation, lead to non-homogeneous surface tension that generates imbalance in forces acting on the floating object.
Both asymmetric and symmetric systems powered by dissipated camphor molecules were studied.
A plastic boat with a camphor piece at its rear, when it is placed on a water surface, moves towards its front, because the surface tension close to the camphor piece is lower~\cite{Kohira,Shimokawa}.
A perfectly circular camphor disk floating on water, can start to move in a certain direction that is determined by a fluctuation in camphor surface concentration~\cite{Chen2009,Koyano2017,Koyano2019}.
In many cases the time evolution of both asymmetric and symmetric camphor-propelled objects can be qualitatively reproduced with a mathematical model that combines a reaction-diffusion equation for the camphor release with a Newtonian equation for the motion of objects\cite{HayashimaJPCB,NagayamaPhysD2004,NakataBook}.
In this model, hydrodynamic flows are described by an effective diffusion constant in the reaction-diffusion equation~\cite{SuematsuLangmuir,KitahataJCP,Bickel2019}.

Different types of evolution of moving camphor-propelled objects have been analyzed in terms of the bifurcation theory~\cite{HayashimaJPCB,NakataBook,NagayamaPhysD2004,Koyano2016,Koyano2019}.
For example, in a recent study we considered a self-propelled rotor composed of a plastic plate with camphor pills at its edges~\cite{Koyano2017,KoyanoChaos}.
The stable rotation is supported by the feedback between two process: the rotational motion induces the asymmetry in the camphor surface concentration, which generates the torque that drives the rotor.
By changing the friction coefficient, the rest state of the rotor becomes unstable and either clockwise or counterclockwise rotation appears.
This change in the character of evolution can be understood as a pitchfork bifurcation.
In the experiments on such systems~\cite{Koyano2017,KoyanoChaos}, the rotor was elevated above the water surface such that only the camphor pills had contact with water.
This was done intentionally to reduce the effect of hydrodynamics and make the model mentioned above more appropriate for description of the observed phenomena.

\begin{figure}[tb]
\begin{center}
\includegraphics{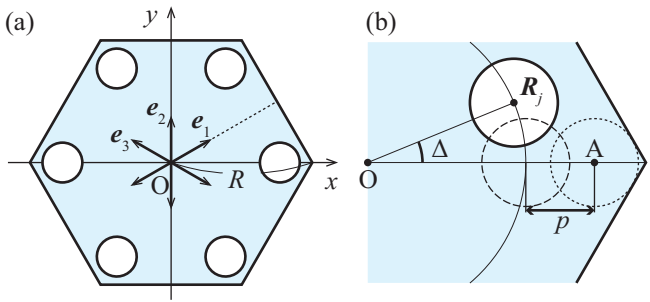}
\end{center}
\caption{(a) A schematic top view of the hexagonal rotor considered in numerical simulations with the marked positions of camphor pills (white disks).
 The unit vectors defined in Eq.~\eqref{ejt0} are also shown.
(b) The magnified illustration of pill position at a corner of the hexagonal plate. The centers of camphor pills with the radius $\rho$ are attached at the points $\bm{R}_j$, which are defined in Eq.~\eqref{rj}. The pill positions are slightly rotated from the line connecting the center and the polygon corner, where the angular shift is $\Delta$. The distances of all pills from the rotor center are $R_c$ (see Eq.~\eqref{rc}). For the pill tangential to the both corner sides (located at the point~A), the distance between A and the center would be $R - \rho / \cos(\pi / N)$. The positive value $p$ is introduced to ensure that, for the used values of $\Delta$, the shifted camphor pills are always inside the hexagonal plate.}
\label{fig1}
\end{figure}

\begin{figure}[tb]
\begin{center}
\includegraphics{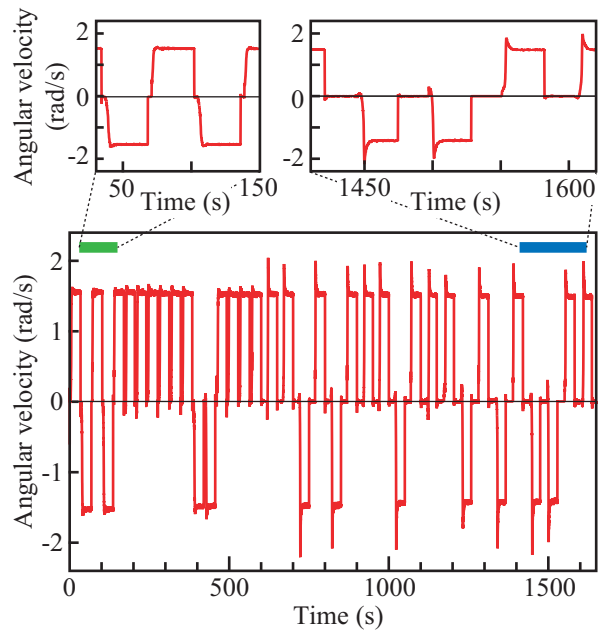}
\end{center}
\caption{Experimental demonstration on the angular velocity as a function of time for a plastic hexagonal plate with the side length 25~mm propelled by 6 camphor pills with the diameter 3~mm each. The upper subfigures expand the angular velocity observed in the time intervals [30 s, 150 s] (stop phase duration: 3 s) and [1410 s, 1620 s] (stop phase duration: 20 s), which are indicated by green and blue bars.}
\label{fig2}
\end{figure}

Reports on a new type of interesting non-equilibrium evolutions of camphor-driven systems, that combine camphor dissipation with hydrodynamics and require a more complex model for their description, have been recently published. For example, we have studied rotation of a hexagonal plate propelled by camphor pills located at its corners (cf. Fig.~\ref{fig1}) under periodic stop-and-release operations~\cite{NakataJPCC}.
In these experiments, the vertical rotation axis was fixed at the hexagon center so that the rotor exhibited only rotational motion.
It has been observed that if such a rotor was stopped for a few seconds then, after its release, the rotational direction was inverted.
When the stop phase duration increased, the memory on the sign of the angular velocity before the stop was lost and the probability of inversion decreased.
Experimental results for the angular velocity of a hexagon are shown in Fig.~\ref{fig2}. The release time was fixed at 30~s.
For such a hexagonal plate, the angular velocity approached its stationary value ($\sim$1.6~rad/s) within less than 10~s, and thus we believe the system was in the stationary state when the stop operation was applied.
The stop phase duration was gradually increased during the experiment.
If it was equal to 3~s (indicated by a green bar, the time interval [30~s, 150~s]), we observed the perfect inversion.
When it was increased to 20~s (the blue bar, the time interval [1410~s, 1620~s]), the angular velocity after stops were selected randomly.
The effect can be explained by the coupling between the camphor surface concentration and the induced surface water flow.
When the rotor is stopped, the surrounding water flow continues for some time and transports camphor molecules around the rotor.
Therefore, the camphor surface concentration around the hexagon corners changes.
When the rotor is released again, the torque resulting from gradients of the surface tension reverts and the inversion of rotational direction is observed. 
The memory about rotational direction before the stop, is preserved in the surrounding hydrodynamic flows and it decays in time. Therefore, the probability of inversion in rotational direction depends on the time duration of the stop phase.
As it may be expected, the maximum duration of the stop phase for which the inversion is observed after every stop depends on the rotor geometry.

As mentioned above, the simple model of evolution for camphor-propelled objects, that treats Marangoni flows by an effective diffusion constant, can qualitatively describe many observed phenomena~\cite{SuematsuLangmuir,KitahataJCP,Bickel2019}.
However, this model is not applicable to the inversion of rotational direction under stop-and-release operations, because it does not treat separately the evolution of flows and the camphor surface concentration.
In the present paper, we propose a new mathematical model for camphor-driven rotor that includes separate time evolution of the flow field.
In order to show its applicability, we have performed numerical calculations for the time evolution of a camphor-driven hexagonal rotor under periodic stop-and-release operations.
We obtained a qualitative agreement with experiments.
Moreover, we discuss the effect of configuration asymmetry, which cannot be avoided in the real experiments.

\section{Model} 

Let us consider a rotor composed of a plate and camphor pills that generate its motion.
The plate has the shape of a regular polygon with $N$ corners, at which $N$ camphor pills are attached.
The center of the plate is fixed at the origin in a two-dimensional coordinate system.
The radius of the circumscribed circle of the polygon is $R$. Let $\bm{e}(\varphi)$ denote the unit vector in the direction of $\varphi$: 
\begin{equation}
\bm{e}(\varphi) = \left( \begin{array}{c} \cos\varphi \\ \sin\varphi \end{array}\right).
\end{equation}
We assume that initially the polygon is oriented such that unit vectors directing the midpoints of polygon edges $\bm{e}_j$ $(j = 1, 2, \cdots, N)$ are: 
\begin{equation}
\bm{e}_j(t=0) = \bm{e}\left( \frac{2 \pi}{N}\left( j - \frac{1}{2}\right) \right). \label{ejt0}
\end{equation}
The region inside the regular polygon is the one of whose corners is on the positive part of the $x$-axis, so the initial position of the polygon can be defined as:
\begin{equation}
\Omega_0 = \left\{ \bm{r} \middle | \min_{j} \left(R \cos \left(\frac{\pi}{N}\right) - \bm{r} \cdot \bm{e}_j(t=0) \right) \geq 0 \right\}.
\end{equation} 

The time evolution of the rotor is defined by time dependent angle $\varphi(t)$. 
Therefore, the region below the plate can be described as:
\begin{equation}
\Omega(\varphi(t)) = \left\{\mathcal{R}(\varphi(t)) \bm{r} \mid \bm{r} \in \Omega_0\right\},
\end{equation}
where $\mathcal{R}(\varphi)$ is the rotation matrix
\begin{equation}
\mathcal{R}(\varphi) = \left( \begin{array}{cc} \cos \varphi & -\sin \varphi \\ \sin \varphi & \cos \varphi \end{array} \right).
\end{equation}

The camphor pills are located at the distance $R_c$ from the center:
\begin{equation}
R_c = R - \frac{\rho}{\cos\left(\pi/N\right)} - p, \label{rc}
\end{equation}
where $\rho$ is the radius of the camphor pills.
Here the first two terms on the right side represent the distance of a pill that is tangential to the both corner sides (located at the point A in Fig.~\ref{fig1}b) from the origin.
The parameter $p \ge 0$ allows to introduce asymmetry in pill location without moving the pill outside the plate. We set the pill positions at:
$\bm{R}_j(t)$ ($j = 1, 2, \cdots , N$; cf. Fig.~\ref{fig1}b), which are defined as:
\begin{equation}
\bm{R}_j(t) = R_c \bm{e} \left( \varphi(t) + \frac{2 \pi}{N} \left ( j - \frac{1}{2} \right ) + \Delta \right). \label{rj}
\end{equation}
The parameter $\Delta$, which denotes the angular shift of the camphor pill positions from the lines connecting the origin and the corners of the regular polygon, introduces asymmetry to the problem.

For the numerical simulations, we define smoothed functions $F(\bm{r}, \varphi(t))$ and $G(\bm{r}, \varphi(t))$ describing the region below the hexagonal plate as:
\begin{align}
&F(\bm{r}, \varphi(t)) \nonumber \\ & =\frac{1}{2} \left[ 1 + \tanh \left( \frac{1}{\delta} \left\{ \min_{j} \left(R \cos \left(\frac{\pi}{N}\right) - \bm{r} \cdot \bm{e}_j(t) \right) \right\} \right) \right],
\end{align}
and the camphor pills as:
\begin{equation}
G(\bm{r}, \varphi(t)) = \sum_{j=1}^N \frac{1}{2} \left[ 1 + \tanh \left( \frac{1}{\delta} \left( \rho - \left| \bm{r} - \bm{R}_j(t) \right| \right)\right)\right]. 
\end{equation}
For simplicity, the same smoothing parameter $\delta$ is used in both $F(\bm{r}, \varphi(t))$ and $G(\bm{r}, \varphi(t))$ functions. 

Our mathematical model for the system evolution combines the motion of the camphor driven rotor described by $\varphi(t)$ with time dependent 
 camphor surface concentration $c(\bm{r},t) = c(x,y,t)$, and the surface flow field $\bm{v}(\bm{r},t) = v_x(x,y,t) \bm{e}_x + v_y(x,y,t) \bm{e}_y$.
The time evolution of $c(\bm{r},t)$ is described as:
\begin{align}
&\frac{\partial c(\bm{r},t)}{\partial t} + \nabla \cdot \left( c(\bm{r},t) \bm{v} (\bm{r},t) \right) = \nonumber \\
&\quad D \nabla^2 c(\bm{r},t) - a \left[ 1- F(\bm{r}, \varphi(t)) \right] c(\bm{r},t) + \frac{s_0}{\pi \rho^2} G(\bm{r} ,\varphi(t)), \label{eqc}
\end{align}
where $D$ is the diffusion coefficient, $a$ is the evaporation rate of camphor molecules to the air phase, and $s_0$ is the supply rate of camphor molecules from a single pill.

The evolution of the flow field $\bm{v}(\bm{r},t)$ can be calculated using:
\begin{align}
\frac{\partial \bm{v}{(\bm{r},t)}}{\partial t} =& \eta F(\bm{r}, \varphi(t)) \left[ \frac{d \varphi}{dt} \left( \begin{array}{c} -y \\ x \end{array} \right) 
 - \bm{v}{(\bm{r},t)} \right] \nonumber \\
& - \eta_b \bm{v}{(\bm{r},t)} + D_\eta \nabla^2 \bm{v}{(\bm{r},t)}. \label{eqv}
\end{align}
The first term of the righthand side describes flow relaxation to the velocity of the hexagonal plate.
The second and third terms represent the viscous effect, i.e., the second term is the viscous effect in the vertical direction, and the third term is in the horizontal plane. 

The time evolution of the hexagonal plate is given by:
\begin{align}
I\frac{d^2\varphi}{dt^2} =& 
- \mu \int_{\Omega(\varphi(t))} \left[(\bm{r}' \cdot \bm{r}') \frac{d\varphi}{dt} - \bm{r}' \times \bm{v}(\bm{r}',t) \right] d\bm{r}'\nonumber \\ 
&+ \mathcal{T}(t) + \xi(t), \label{eq_motion}
\end{align}
where $I$ is the moment of inertia, $\mu$ is the friction constant, and $\mathcal{T}$ is the torque, which is described as:
\begin{equation}
\mathcal{T}(t) = \int_{\Omega(\varphi(t))} \bm{r}' \times \left( - \nabla \gamma(c(\bm{r}',t)) \right) d\bm{r}'. \label{torque}
\end{equation}
$\gamma(c)$ is the surface tension which is a decreasing function of $c$ as reported in the previous work\cite{camphor_surface_tension}. Here, we assume
\begin{equation}
\gamma(c) = \gamma_0 - k c, \label{surfacetension}
\end{equation}
where $\gamma_0$ is the surface tension of water, and $k$ is a positive constant.
To include stochasticity into the model, we have introduced a white Gaussian noise term $\xi(t)$ whose average and standard deviation are 0 and $\sigma$, respectively:
\begin{equation}
\left<\xi(t) \right> = 0,
\end{equation}
\begin{equation}
\left< \xi(t) \xi(s) \right> = 2 \sigma^2 \delta(t - s). \label{noiseav}
\end{equation}
We adopted Box-Muller's method \cite{num-rec} to generate the white Gaussian noise. 

For the following analysis we use the initial conditions:
\begin{align}
&\varphi(t=0) = \varphi_0, \nonumber \\
&\frac{d\varphi}{dt}(t=0) = \omega_0, \nonumber \\
&\bm{v}{(\bm{r},t=0)} \equiv \mathbf{0}, \nonumber \\
&c{(\bm{r},t=0)} \equiv 0.
\end{align}

After simulations are initiated, we follow the evolution of the system for the time $t_\textrm{ini}$ to achieve the stationary rotation.
Next, we introduce the stop-and-release operations.
Let us assume that the rotor is forced to stop for the time interval $t_\textrm{stop}$ and it is released for the time interval $t_\textrm{release}$. Such stop-and-release operations are repeated for $M$ times.
The period of a single stop-and-release cycle is $T=t_\textrm{stop}+t_\textrm{release}$.
Let us also assume that the first stop starts at $t = t_\textrm{ini}$. In order to
introduce the forcing stop into equations~\eqref{eq_motion}, we assume that:
\begin{equation}
 \varphi(t) = \varphi(t_\textrm{ini} +(m-1) T) 
\end{equation}
for $t \in (t_\textrm{ini} +(m-1) T, t_\textrm{ini} +t_\textrm{stop} +(m-1) T]$ and $m=1,\cdots ,M$. Therefore,
\begin{equation}
\frac{d\varphi(t)}{dt} = 0,
\end{equation}
if $t \in (t_\textrm{ini} +(m-1) T, t_\textrm{ini} +t_\textrm{stop} +(m-1) T)$. 
The conditions for the camphor surface concentration $c{(\bm{r},t)}$ in Eq.~\eqref{eqc} and for the flow field $\bm{v}{(\bm{r},t)}$ in Eq.~\eqref{eqv} are not modified.

\section{Numerical results}

In numerical calculations, we considered a regular hexagonal plate, i.e., $N=6$.
The values of $t_\mathrm{stop}$ and $\Delta$ were regarded as the control parameters.
For all simulations $t_{\mathrm{ini}} = t_{\mathrm{release}} = 10$.
The number of the stop- and release- cycles was $M=100$.
The diffusion constant $D$ and the evaporation rate $a$ of the concentration field were set to be 1 for each. Thus, the diffusion length of the concentration field $\sqrt{D/a}$ was also 1. In the experiments, the diffusion length was several tens of millimeters, and the sides of the hexagonal plate and pill diameter were 25 mm and 3 mm, respectively. Thus, the values of $R (=0.5)$, $\rho (=0.1)$, and $p (= 0.05)$ were realistic.
The moment of inertia was set to be $I = 10^{-5}$ so that the relaxation time of the angular velocity is fast enough compared with the $t_{\mathrm{release}}$.
The values of $s_0$ and $k$, which determine the magnitude of the driving force of the plate, were set to be 1 for each in an arbitrary way. Then the friction coefficient was set to be $\mu = 10^{-4}$, so that the hexagonal rotor does not show a rest state but a steady rotation.
The smoothing parameter $\delta$ was set to be $\delta = 0.025$, which is equal to the spatial mesh size to avoid the effect of the mesh.
The parameters $\eta$, $\eta_b$, and $D_\eta$, appearing in Eq.~\eqref{eqv}, were all set to be 1, which indicates that the characteristic temporal and spatial decay scales of the flow field are both 1.
With such parameter values, the stationary angular velocity of the rotor was $\sim 176$ and it was almost independent of $\Delta$ if $\left| \Delta \right| \lesssim 10^{-4}$.
The characteristic relaxation time of the angular velocity was in the order of 1, which was 10 times shorter than $t_{\mathrm{ini}}$ and $t_{\mathrm{release}}$.

We adopted the explicit method for the spatial derivative, and used the Euler method for the time evolution.
The spatial and time steps were: $\Delta x = 0.025$ and $\Delta t = 10^{-4}$, respectively.
The calculation field was a circular region with a radius of 3, i.e., 3 times as large as the diffusion length and 6 times larger than $R$. The results shown in Fig.~\ref{fig2} were obtained for $R=2.5$ cm and the rotor was moving inside a Petri dish with the diameter $18$ cm. The noise amplitude was set to be $\sigma = 0.1$.
If the absolute value of generated noise $\xi$ was greater than 10, then such a value was not used in a numerical step to avoid the divergence of numerical calculation.

The Dirichlet boundary conditions were adopted for both camphor surface concentration and for the surface flow field, i.e., $c{(\bm{r},t)}=0$ and $\bm{v}{(\bm{r},t)} = \mathbf{0}$ at the boundary.
The initial conditions were $\varphi_0 = 1$ and $\omega_0 = 0.1$. 

\begin{figure}[tb]
\begin{center}
\includegraphics{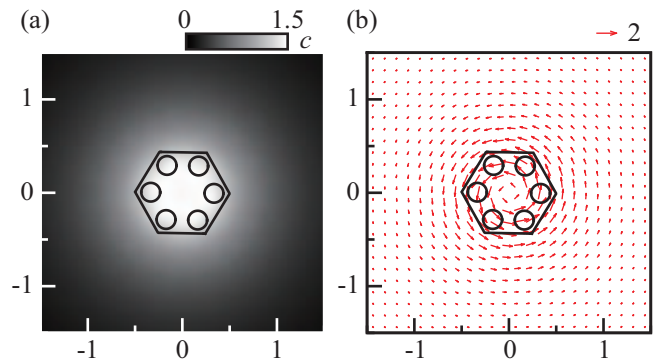}
\end{center}
\caption{The camphor surface concentration (a) and the flow field (b) for a stationary rotating plate in a counterclockwise direction. The rotor is symmetric ($\Delta = 0$). The snapshots show the central area of the computing grid, represented by a square with the side length of 3. The color bar (a) and the scale of the vector (b) are shown above each panel.}
\label{fig_c_v}
\end{figure}

Figure~\ref{fig_c_v}(a) shows a snapshot of calculated camphor surface concentration.
The asymmetry in camphor surface concentration originating from the rotational motion of the plate was not clearly observed, because the surface concentration was dominated by locations of the pills.
Figure~\ref{fig_c_v}(b) shows the flow field for a stationary rotating plate in a counterclockwise direction.
The flow profile was concentric around the object, and decayed with the distance from the center. The flow outside the plate continued to transport camphor after the plate was stopped.
The flow reversed the asymmetry in camphor surface concentration around the corners of a stopped plate and, after the plate was released, the resulting torque forced the rotation in the opposite direction.
Thus the present model can predict inversion of the rotor.
The asymmetry in camphor concentration works as a memory, and it is expected to be lost for long stop phase because the flow field disappears.

\begin{figure}[tb]
\begin{center}
\includegraphics{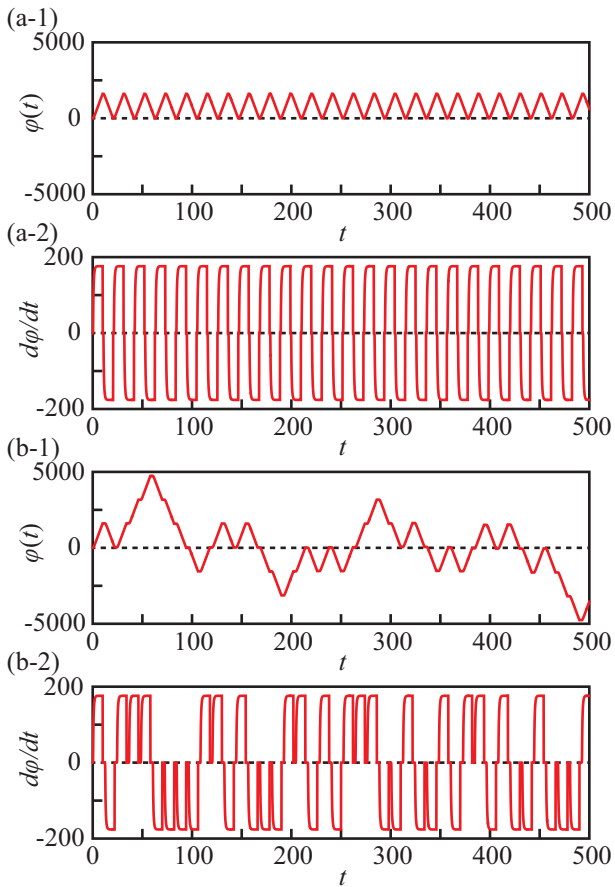}
\end{center}
\caption{Results of numerical simulations for the symmetric rotor ($\Delta = 0$). (1) The characteristic angle $\varphi(t)$ and (2) the angular velocity $d\varphi/dt$ as functions of time for (a) $t_{\mathrm{stop}} = 0.5$ and (b) $t_{\mathrm{stop}} = 2$.}
\label{fig_sym_ts}
\end{figure}

\begin{figure}[tb]
\begin{center}
\includegraphics{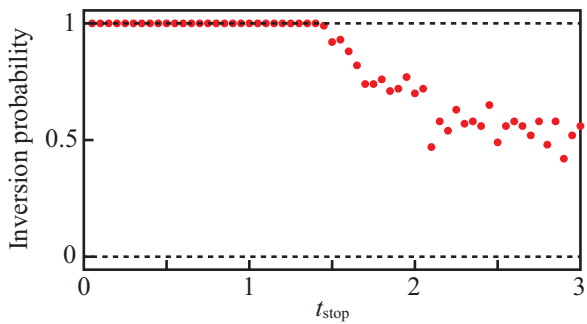}
\end{center}
\caption{Inversion probability as the function of $t_\textrm{stop}$ for the symmetric rotor ($\Delta = 0$).}
\label{fig_sym_phase}
\end{figure}

First, we made numerical simulation for the symmetric rotor ($\Delta = 0$), the result of which is shown in Fig,~\ref{fig_sym_ts}.
After every stop phase, the rotor began to rotate clockwise or counterclockwise and reached its stationary angular velocity.
The direction of the rotation was always inverted for $t_{\mathrm{stop}}=0.5$, as shown in Fig.~\ref{fig_sym_ts}(a).
In contrast, for $t_{\mathrm{stop}}=2$ the selected rotational direction was random as shown in Fig.~\ref{fig_sym_ts}(b).
The inversion probability is defined as the number of inversions divided by the total number of stop-and-release operations $M$.
The inversion probability as the function of the stop phase duration $t_{\mathrm{stop}}$ is illustrated in Fig.~\ref{fig_sym_phase}. 
For the smaller $t_{\mathrm{stop}}$, the inversion probability was 1.
It decayed around $t_{\mathrm{stop}} \simeq 1.4$ and converged to 0.5 for greater $t_{\mathrm{stop}}$.

\begin{figure}[tb]
\begin{center}
\includegraphics{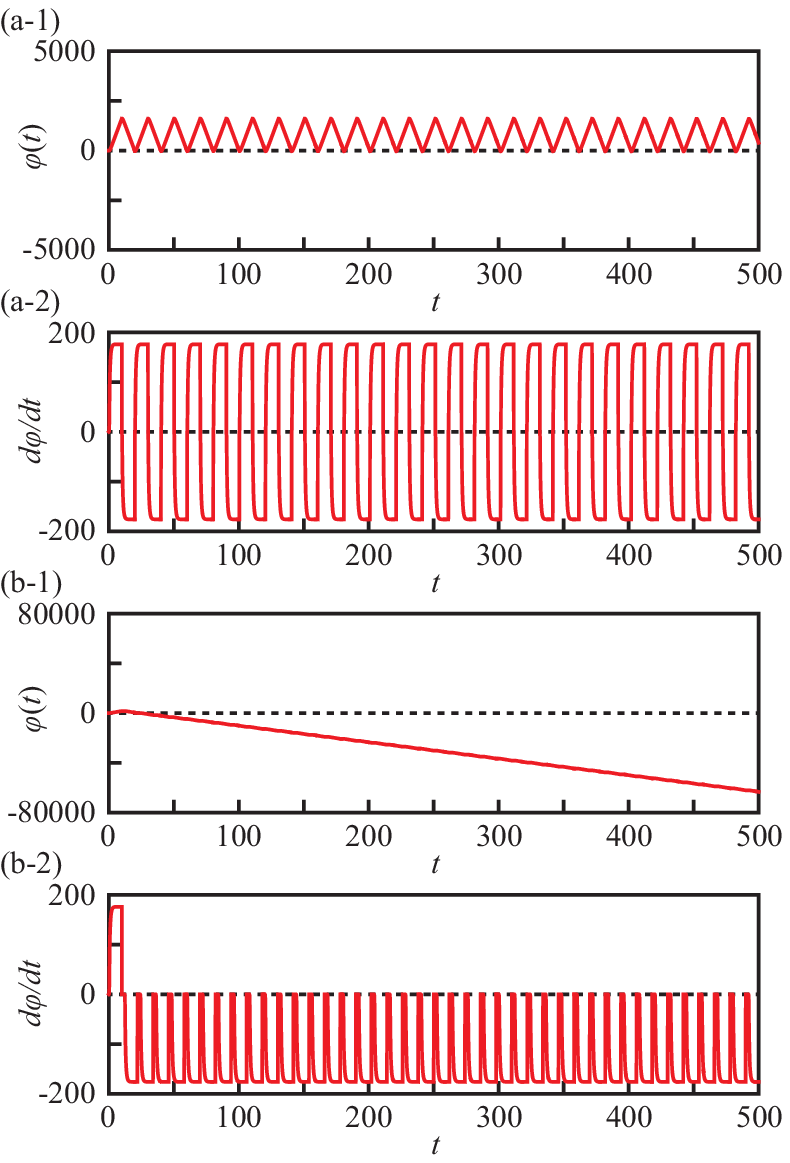}
\end{center}
\caption{Results of numerical simulations for the asymmetric rotor ($\Delta =10^{-4}$). (1) The characteristic angle $\varphi(t)$ and (2) the angular velocity $d\varphi/dt$ as functions of time for (a) $t_{\mathrm{stop}} = 0.5$ and (b) $t_{\mathrm{stop}} = 2$.}
\label{fig_asym_ts}
\end{figure}

\begin{figure}[tb]
\begin{center}
\includegraphics{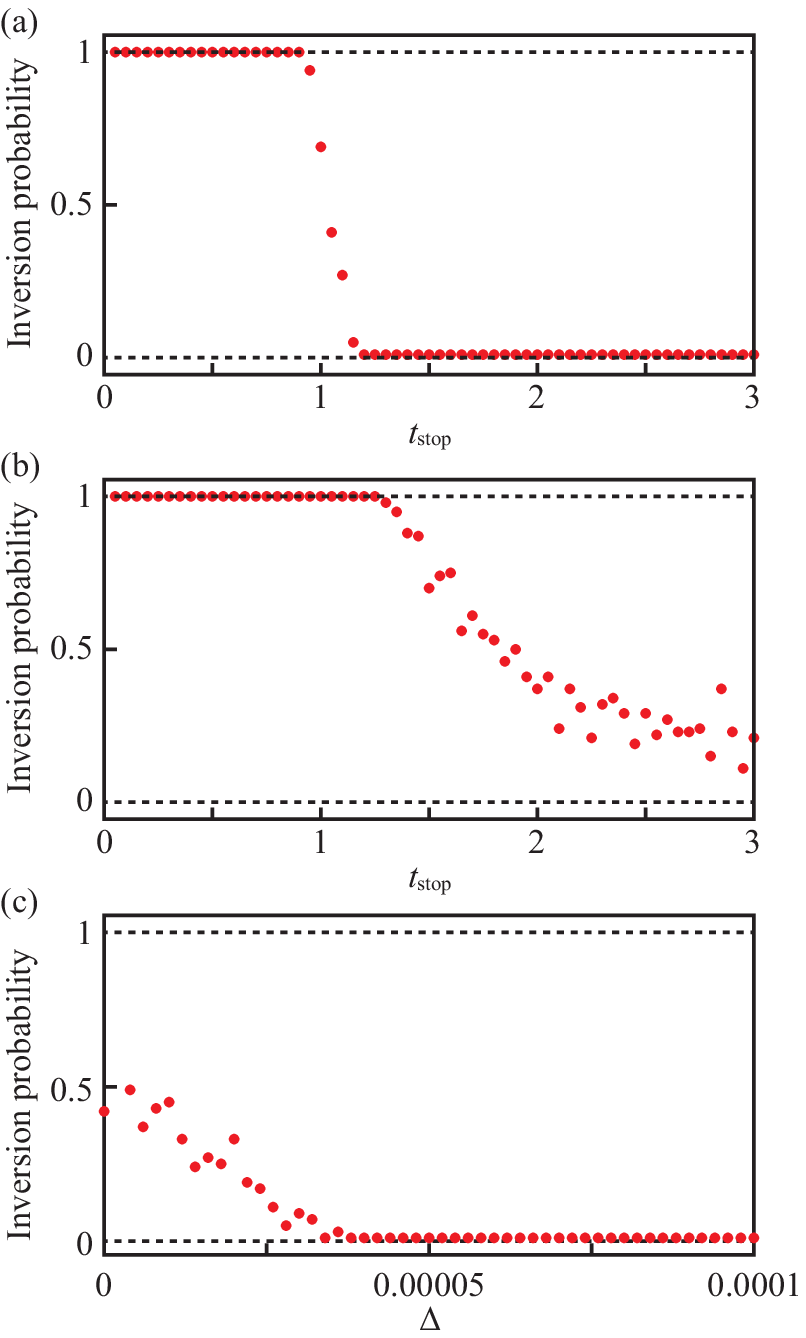}
\end{center}
\caption{The inversion probability as the function of $t_\textrm{stop}$ for the asymmetric rotors. (a) $\Delta = 10^{-4}$, (b) $\Delta = 2 \times 10^{-5}$. (c) The inversion probability as the function of the asymmetry parameter $\Delta$ for long time duration of stop phase, $t_\mathrm{stop} = 3$.}
\label{fig7}
\end{figure}

Next, we investigated the case of an asymmetric rotor, which is represented by a finite $\Delta$ with relatively small absolute value.
Such a rotor has a preferred rotational direction determined by $\Delta$, i.e. it tended to rotate clockwise for $\Delta > 0$ and counterclockwise for $\Delta < 0$.
Nevertheless, the perfect inversion in the rotational direction was observed for $t_\mathrm{stop}=0.5$, as shown in Fig.~\ref{fig_asym_ts}(a).
On the contrary, for $t_{\mathrm{stop}}=2$ the rotor did not exhibit the inversion in the rotational direction, but always rotated in its pre-determined preferable direction, as illustrated in Fig.~\ref{fig_asym_ts}(b).
In Fig.~\ref{fig7}(a), the inversion probability as a function of the stop phase duration $t_{\mathrm{stop}}$ for $\Delta = 10^{-4}$ is shown. 
For small $t_{\mathrm{stop}}$, the inversion probability was 1 just as it was $\Delta = 0$.
It started to decay around $t_{\mathrm{stop}} \simeq 1$ and converged to 0 for longer $t_{\mathrm{stop}}$, which was different than in the symmetric case.
This means that for long $t_{\mathrm{stop}}$ the asymmetric rotor always selected the favorable rotational direction.

\begin{figure}[tb]
\includegraphics{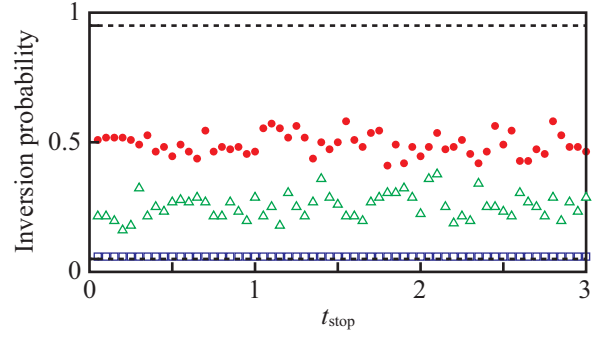}
\caption{Inversion probability in the case with no flow effect as the function of $t_\textrm{stop}$ for the symmetric (red closed circles: $\Delta = 0$)
and asymmetric (green open triangles: $\Delta =2 \times 10^{-5}$, blue open squares: $\Delta =10^{-4}$) rotors. To obtain these results, we neglected Eq.~\eqref{eqv} and set $\bm{v}{(\bm{r},t)} \equiv \bm{0}$ in Eqs.~\eqref{eqc} and \eqref{eq_motion}.}
\label{fig_nohydro}
\end{figure}

It should be stressed that the inversion in the rotational direction was not observed if the flow effects were neglected, i.e. $\bm{v}{(\bm{r},t)} \equiv \bm{0}$ in Eqs.~\eqref{eqc} and \eqref{eq_motion}. Figure~\ref{fig_nohydro} illustrates the results for symmetric ($\Delta = 0$) and asymmetric ($\Delta =10^{-4}$ and $\Delta =2 \times 10^{-5}$) rotors. In the symmetric case, the system was dominated by noise and the direction of rotation after a forced stop was chosen randomly in all range of $t_{\mathrm{stop}}$. On the other hand, the highly asymmetric rotor ($\Delta =10^{-4}$) always selected the same rotational direction. For a lower value of the asymmetry parameter ($\Delta =2 \times 10^{-5}$), the competition between asymmetry and noise can be seen. Here the favorable rotational direction is selected with probability $0.8$ after a forced stop. Nevertheless, in all cases shown in Fig.~\ref{fig_nohydro}, the systematic inversion of rotational direction was not observed for any $t_{\mathrm{stop}}$.

\section{Discussion and Conclusions}

Using the present model, we confirmed in numerical simulations that the inverse in the rotational direction of the rotor for smaller $t_\mathrm{stop}$ is caused by the asymmetric profile of the camphor surface concentration, resulting from the transport of the camphor molecules by the flow during the stop phase, as it was postulated in the previous study\cite{NakataJPCC}.
The asymmetry in the camphor surface concentration decays in time due to the evaporation.
Thus, for the symmetric case ($\Delta = 0$), the competition between the flow-induced asymmetric camphor surface concentration and the noise takes place.
The asymmetric concentration dominates for smaller $t_{\mathrm{stop}}$, while the effect of noise becomes important for larger $t_{\mathrm{stop}}$.
For the asymmetric case ($\Delta = 10^{-4}$), the competition between the flow-induced asymmetric camphor surface concentration and the intrinsic asymmetry in the positions of pills influences the time evolution.
The asymmetry in camphor surface concentration plays the major role for smaller $t_{\mathrm{stop}}$, whereas the asymmetry of the rotor dominates for larger $t_{\mathrm{stop}}$.

\begin{figure}[tb]
\includegraphics{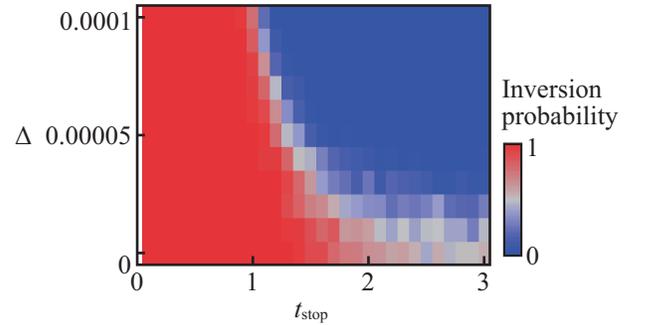}
\caption{Inversion probability as a function of the time duration of stop phase $t_\mathrm{stop}$ and the asymmetry parameter $\Delta$.}
\label{fig_9}
\end{figure}

Using the arguments presented above, we expect that, if the value of the asymmetry parameter $\Delta$ is changed, then the competition between the noise and the intrinsic asymmetry in the rotor should be observed for the larger $t_{\mathrm{stop}}$.
In order to verify it, we calculated the inversion probability as the function of the asymmetry parameter $\Delta$ for a long stop phase ($t_{\mathrm{stop}}=3$). The results are shown in Fig.~\ref{fig7} (c).
As expected, the inversion probability decreased from $0.5$ to $0$ with an increase in $\Delta$.
It indicates that the behavior of the rotor changed from the one dominated by the external noise to the one dominated by inherent rotor asymmetry.
As expected, the inversion probability decreases from 0.5 to 0 with an increase in $\Delta$. It indicates that the behavior of the rotor changes from the one dominated by the external noise to the one dominated by inherent rotor asymmetry. The inversion probability as a function of the time duration of stop phase $t_\mathrm{stop}$ and the asymmetry parameter $\Delta$ is shown in Fig.~\ref{fig_9}. In the red parameter region, the probability of rotation inversion is high, since the effects of the noise and the intrinsic asymmetry are relatively small. In the blue region, the effect of the intrinsic asymmetry is dominant, and the rotor tends to rotate in a preferable direction determined by the intrinsic asymmetry. The gray region around the longer $t_\mathrm{stop}$ and smaller $\Delta$, the noise dominates and thus the rotational direction becomes random.

In the reported experimental results\cite{NakataJPCC}, the inversion probability converged to the finite value around 0.2 for larger $t_{\mathrm{stop}}$.
This value reflects the balance between the system asymmetry due to experimental errors in preparing the rotor and the time-dependent noise in the concentration field and the flow field. We have performed the numerical calculation for the inversion probability as a function of $t_\mathrm{stop}$ for a smaller value of the asymmetry parameter $\Delta = 2 \times 10^{-5}$. For this value of $\Delta$, the inversion probability as the function of $t_{\mathrm{stop}}$, illustrated in Fig.~\ref{fig7} (b), looks similar to the one observed in experiments\cite{NakataJPCC}.

Complete mathematical description of objects propelled at the water surface by the dissipation of camphor is expected to be complex. It should take into account the spread of camphor on the surface, its evaporation and dissipation. The spread of camphor depends on the advection generated by the flow field and on the location of camphor sources~\cite{PCCP,Ikura,CSA}. The flow field is described by a hydrodynamic equation with the source term resulting from the motion of objects and the boundary condition to include the effect of surface tension gradient. This equation should also take water incompressibility into account. Experimental results indicate that a complex 3-dimensional flow may appear around a camphor source. Finally, the surface tension gradients, which are directly related to local values of camphor surface concentration, generate forces and torques propelling objects on the surface and changing positions of camphor sources. Therefore, a mathematical model of time evolution for such systems should combine equations for 2-dimensional surface camphor concentration with hydrodynamics of 3-dimensional flows and Newtonian equations for the object motion~\cite{Lauga}. The solution of these equations should be numerically complex, especially when the geometry of moving objects is not simple. In order to obtain a working model capable to explain experimental results, simulations should be simplified.

The most popular model used to describe time evolution of objects propelled by camphor particles neglects the flow field and characterizes the system with two variables: the camphor surface concentration and the location of objects on the water surface~\cite{NakataBook,Koyano2019,HayashimaJPCB,NagayamaPhysD2004,Koyano2016}. In such models, the Newtonian equation describes object motion and a reaction-diffusion equation with an effective value of diffusion constant, that incorporates advection, represents the time evolution of camphor surface concentration. Such a model has been successfully used for qualitative explanation of many experimental results~\cite{SuematsuLangmuir,KitahataJCP,Bickel2019}. However, it cannot be applied for the case when the object motion is externally perturbed and the flow evolution is not determined by location of objects and by camphor surface concentration. For example, it happens when forced stop-and-release operations are applied to the moving objects.
Moreover, the considered system is one of examples of unconventional motors\cite{NakataBook}, where the dynamics of the angle of the motor, the concentration field, and the flow field cooperatively realize a function. In this sense, our results allow to check if an unconventional motor can operate with the required level of accuracy under the intrinsic inaccuracy in the system geometry and the present spatio-temporal random noise.

In this paper, we have presented a new simple model for the time evolution of self-propelled objects on a water surface that describes the system using three independent quantities: the object location, the surface concentration of active molecules and the hydrodynamic flows on the surface. The equation for the time evolution of surface concentration includes advection, but still the spread resulting from Marangoni flow is described by an effective diffusion coefficient. Our equation for the time evolution of the flow field has just one source term linking the flow with the object motion. The numerical complexity of such model remains relatively low and simulations can be performed on a personal computer. In our opinion, the new model represents a significant improvement if compared with the two-variable one, because it incorporates the flows directly to evolution equations. Using the new model, we succeeded in reproducing the experimental results\cite{NakataJPCC}, in which the rotor exhibits the inversion of the rotational direction under periodic stop-and-release operations.
The model also correctly reproduced the probability of the inversion as a function of the duration of the stop phase when an asymmetry was introduced into the rotor configuration, which is unavoidable in the process of preparing the rotor in the experiments.
The model confirmed that the inversion probability of the rotor rotational direction is determined by the competition among the three factors: transport of the camphor molecules by the flow, the intrinsic asymmetry of the rotor, and the noise amplitude. 
The present work exemplifies how the required function (here it is the inversion of the rotation) is affected by the systematic error of the system (the asymmetric parameter) and the spatio-temporal random noise (noise intensity). 
We believe the developed method can be used to describe other systems where motion of self-propelled objects on a liquid surface is strongly coupled with flow dynamics. Such systems include those composed of individually moving self-propelled objects, where complex modes of evolution are observed.

\begin{acknowledgments}
The authors are grateful to Dr. Marian Gryciuk for his help in processing experimental results shown in Fig.~\ref{fig2}. This work was supported by a Bilateral Joint Research Program ``Spatio-temporal patterns of elements driven by self-generated, geometrically constrained flows'' between Japan and the Polish Academy of Sciences and by JSPS KAKENHI Grant Numbers JP16H03949, JP19J00365, 17K05835,
and 17KT0123, and the Cooperative Research Program of ``Network Joint Research Center for Materials and Devices'' Nos.~20191030 and 20194006.

\end{acknowledgments}


\begin{thebibliography}{99}

\bibitem{NakataJPCC} S.~Nakata, K.~Kayahara, H.~Yamamoto, P.~Skrobanska, J.~Gorecki, A.~Awazu, H.~Nishimori, and H.~Kitahata, \textit{J. Phys. Chem. C} \textbf{122}, 3482 (2018).

\bibitem{Ramaswamy2010} S.~Ramaswamy, \textit{Annu. Rev. Cond. Mat. Phys.} \textbf{1}, 323 (2010).

\bibitem{Marchetti2013} M.~C.~Marchetti, J.~F.~Joanny, S.~Ramaswamy, T.~B.~Liverpool, J.~Prost, M.~Rao, and R.~A.~Simha, \textit{Rev. Mod. Phys.} \textbf{85}, 1143 (2013).

\bibitem{Bechinger2016} C.~Bechinger, R.~Di~Leonardo, H.~L\"{o}wen, C.~Reichhardt, G.~Volpe, G.~Volpe, \textit{Rev. Mod. Phys.} \textbf{88}, 045006 (2016).

\bibitem{Pimienta2014} V.~Pimienta and C.~Antoine, \textit{Curr. Opin. Colloid Interface Sci.} \textbf{19}, 290 (2014).

\bibitem{Keren} K.~Keren, Z.~Pincus, G.~M.~Allen, E.~L.~Barnhart, G.~Marriott, A.~Mogilner, and J.~A.~Theriot, \textit{Nature} \textbf{453}, 475
(2008).

\bibitem{Taylor2013} T.~Ba\'{n}sa\'{g}i, Jr., M.~M.~Wrobel, S.~K.~Scott, and A.~F.~Taylor, \textit{J. Phys. Chem. B} \textbf{117}, 13572 (2013).

\bibitem{Park2017} J.~H.~Park, S.~Lach, K.~Polev, S.~Granick, and B.~A.~Grzybowski, \textit{ACS Nano} \textbf{11}, 10914 (2017).

\bibitem{Frenkel2017} M.~Frenkel, V.~Multanen, R.~Grynyov, A.~Musin, Y.~Bormashenko, and E.~Bormashenko, \textit{Sci. Rep.} \textbf{7}, 3930 (2017).

\bibitem{Nagai2005} K.~Nagai, Y.~Sumino, H.~Kitahata, and K.~Yoshikawa, \textit{Phys. Rev. E} \textbf{71}, 065301 (2005).

\bibitem{Hanczyc2014} J.~\v{C}ejkov\'{a}, M.~Nov\'{a}k, F.~\v{S}t\v{e}p\'{a}nek, and M.~M.~Hanczyc, \textit{Langmuir} \textbf{30}, 11937 (2014).

\bibitem{Lagzi2010} I.~Lagzi, S.~Soh, P.~J.~Wesson, K.~P.~Browne, and B.~A.~Grzybowski, \textit{J. Am. Chem. Soc.} \textbf{132}, 1198 (2010).

\bibitem{Takanatake2014} F.~Takabatake, N.~Magome, M.~Ichikawa, and K.~Yoshikawa, \textit{J. Chem. Phys.} \textbf{134}, 114704 (2011).

\bibitem{Toyota2009} T.~Toyota, N.~Maru, M.~M.~Hanczyc, T.~Ikegami, and T.~Sugawara, \textit{J. Am. Chem. Soc.} \textbf{131}, 5012 (2009).

\bibitem{Izri2014} Z.~Izri, M.~N.~van der Linden, S.~Michelin, and O.~Dauchot, \textit{Phys. Rev. Lett.} \textbf{113}, 248302 (2014).

\bibitem{Sumino2005} Y.~Sumino, N.~Magome, T.~Hamada, and K.~Yoshikawa, \textit{Phys. Rev. Lett.} \textbf{94}, 068301 (2005).

\bibitem{Cira2015} N.~J.~Cira, A.~Benusiglio, and M.~Prakash, \textit{Nature} \textbf{519}, 446 (2015).

\bibitem{Domingues1995} F.~Domingues dos Santos and T.~Ondar\c{c}uhu, \textit{Phys. Rev. Lett.} \textbf{75}, 2972 (1995).

\bibitem{Ohta2017} T.~Ohta, \textit{J. Phys. Soc. Jpn.} \textbf{86}, 072001 (2017).

\bibitem{OhtaOhkuma} T.~Ohta and T.~Ohkuma, \textit{Phys. Rev. Lett.} \textbf{102}, 154101 (2009).

\bibitem{Ebata1} H.~Ebata and M.~Sano, \textit{Sci. Rep.} \textbf{5}, 8546 (2015).

\bibitem{Ebata2} H.~Ebata, A.~Yamamoto, Y.~Tsuji, S.~Sasaki, K.~Moriyama, T.~Kuboki, and S.~Kidoaki, \textit{Sci. Rep.} \textbf{8}, 5153 (2018).

\bibitem{KoyanoJCP} Y.~Koyano, N.~Yoshinaga, and H.~Kitahata, \textit{J. Chem. Phys.} \textbf{143}, 014117 (2015).

\bibitem{Skey} W.~Skey, \textit{Trans. Proc. R. Soc. New Zealand} \textbf{11}, 473 (1878).

\bibitem{Tomlinson} C.~Tomlinson, \textit{Proc. R. Soc. London} \textbf{11}, 575 (1862).

\bibitem{Rayleigh} L.~Rayleigh, \textit{Proc. R. Soc. London} \textbf{47}, 364 (1889).

\bibitem{NakataLangmuir} S.~Nakata, Y.~Iguchi, S.~Ose, M.~Kuboyama, T.~Ishii, and K.~Yoshikawa, \textit{Langmuir} {\bf 13}, 4454 (1997).

\bibitem{Nakata2015} S.~ Nakata, M.~Nagayama, H.~Kitahata, N.~J.~Suematsu, and T.~Hasegawa, \textit{Phys. Chem. Chem. Phys.} \textbf{17}, 10326 (2015).

\bibitem{NakataBook} S.~Nakata, V.~Pimienta, I.~Lagzi, H.~Kitahata, and N.~J.~Suematsu, \textit{Self-organized motion: Physicochemical design based on nonlinear dynamics} (R. Soc. Chem., Cambridge, 2019).

\bibitem{camphor_surface_tension} C.~C.~de~Wit and R.~F.~Makens, \textit{J. Am. Chem. Soc.} \textbf{54}, 455 (1932). 

\bibitem{Karasawa} Y.~Karasawa, S.~Oshima, T.~Nomoto, T.~Toyota, and M.~Fujinami, {\it Chem. Lett.} {\bf 43}, 1002 (2014).

\bibitem{Kohira} S.~Nakata, M.~I.~Kohira and Y.~Hayashima, \textit{Chem. Phys. Lett.} \textbf{322}, 419 (2000).

\bibitem{Shimokawa} M.~Shimokawa, M.~Oho, K.~Tokuda, and H.~Kitahata, \textit{Phys. Rev. E} \textbf{98}, 022606 (2018)

\bibitem{Chen2009} X.~Chen, S.-I. Ei, and M. Mimura, \textit{Netw. Heterog. Media} \textbf{4}, 1 (2009).

\bibitem{Koyano2017} Y.~Koyano, M.~Gryciuk, P.~Skrobanska, M.~Malecki, Y.~Sumino, H.~Kitahata, and J.~Gorecki, \textit{Phys. Rev. E} \textbf{96}, 012609 (2017).

\bibitem{Koyano2019} Y.~Koyano, N.~J.Suematsu, and H.~Kitahata, \textit{Phys. Rev. E} \textbf{99}, 022211 (2019).

\bibitem{HayashimaJPCB} Y.~Hayashima, M.~Nagayama, and S.~Nakata, \textit{J. Phys. Chem. B} \textbf{105}, 5353 (2001).

\bibitem{NagayamaPhysD2004} M.~Nagayama, S.~Nakata, Y.~Doi, and Y.~Hayashima, \textit{Physica D} \textbf{194}, 151 (2004).

\bibitem{SuematsuLangmuir} N.~J.~Suematsu, T.~Sasaki, S.~Nakata, and H.~Kitahata, \textit{Langmuir} \textbf{30}, 8101 (2014).

\bibitem{KitahataJCP} H.~Kitahata and N.~Yoshinaga, \textit{J. Chem. Phys.} \textbf{148}, 134906 (2018).

\bibitem{Bickel2019} T.~Bickel, \textit{Soft Matter} \textbf{15}, 3644 (2019).

\bibitem{Koyano2016} Y.~Koyano, T.~Sakurai, and H.~Kitahata, \textit{Phys. Rev. E} \textbf{94}, 042215 (2016).

\bibitem{KoyanoChaos} Y.~Koyano, H.~Kitahata, M.~Gryciuk, N.~Akulich, A.~Gorecka, M.~Malecki, and J.~Gorecki, \textit{Chaos} \textbf{29}, 013125 (2019).

\bibitem{num-rec} W.~H.~Press, S.~A.~Teukolsky, W.~T.~Vetterling, B.~P.~Flannery, \textit{Numerical Recipes in C. The Art of Scientific Computing, 2nd Ed.} (Cambridge Univ. Press, New York, 1992).

\bibitem{PCCP} H.~Kitahata, S.-i.~Hiromatsu, Y.~Doi, S.~Nakata, and M.~R.~Islam, \textit{Phys. Chem. Chem. Phys.} \textbf{6}, 2409 (2004).

\bibitem{Ikura} Y.~Ikura, R.~Tenno, H.~Kitahata, N.~J.~Suematsu and S.~Nakata, \textit{J. Phys. Chem. B} \textbf{116}, 992 (2012).

\bibitem{CSA} H.~Kitahata, H.~Yamamoto, M.~Hata, Y.~S.~Ikura, and S.~Nakata, \textit{Colloids Surf. A} \textbf{520}, 436 (2017).

\bibitem{Lauga} E.~Lauga and A.~M.~J.~Davis, \textit{J. Fluid Mech.} \textbf{705}, 120 (2012).

\end{thebibliography}
\end{document}